\documentclass[final,5p,times,twocolumn]{elsarticle}
\pdfoutput=1

\usepackage{amssymb}
\usepackage{lipsum}
\usepackage[T1]{fontenc}

\usepackage{graphicx}
\usepackage{dcolumn}
\usepackage{bm}
\usepackage{todonotes} 
\usepackage{hyperref}
\hypersetup{
    colorlinks=true,
    linkcolor=blue,
    filecolor=magenta,      
    urlcolor=cyan,
    }

\usepackage[mathlines]{lineno}
\usepackage{multirow}
\usepackage{adjustbox}
\usepackage{units}

\journal{Physics Letters B}

\begin{document}

\title{Bounding elastic photon-photon scattering at $\sqrt s \approx 1$\,MeV using a laser-plasma platform}

\newcommand{\IC}{The John Adams Institute for Accelerator Science, Imperial College London, London, SW7 2AZ, UK}

\newcommand{\UMICH}{Center for Ultrafast Optical Science, University of Michigan, Ann Arbor, MI 48109-2099, USA}

\newcommand{\DESY}{Deutsches Elektronen-Synchrotron DESY, Notkestr. 85, 22607 Hamburg, Germany}

\newcommand{\QUB}
{School of Mathematics and Physics, Queen's University of Belfast, BT7 1NN, Belfast, UK}

\newcommand{\HIJENA}
{Helmholtz Institut Jena, Fr\"{o}belstieg 3, 07743 Jena, Germany}

\newcommand{\JENA}
{Institut f\"{u}r Optik und Quantenelektronik, Friedrich-Schiller-Universit\"{a}t, 07743 Jena, Germany}

\newcommand{\CLF}{Central Laser Facility, STFC Rutherford Appleton Laboratory, Didcot OX11 0QX, UK}

\newcommand{\YORK}{Department of Physics, University of York, York, YO10 5DD, UK}

\newcommand{\OXFORD}{Department of Physics, University of Oxford, Oxford, OX1 3PU, UK}

\newcommand{\AWE}{AWE Aldermaston, Reading RG7 4PR, UK}

\newcommand{\CELIA}{Université de Bordeaux-CNRS-CEA, Centre Lasers Intenses et Applications (CELIA), UMR 5107, F-33405 Talence, France}

\newcommand{\UVa}{Departamento de Física Teórica Atómica y Óptica, Universidad de Valladolid, 47011 Valladolid, Spain}

\newcommand{\LBNL}{Lawrence Berkeley National Laboratory, Berkeley, CA 94720, US}

\newcommand{\LANC}{Physics Department, Lancaster University, Lancaster LA1 4YB, UK}

\newcommand{\CERN}{CERN, Geneva, Switzerland}
\newcommand{\PLYMOUTH}{Centre for Mathematical Sciences, University of Plymouth, PL4 8AA, UK}

\author[imp]{R.~Watt}
\author[imp]{B.~Kettle}
\author[imp]{E.~Gerstmayr}
\author[ply]{B.~King}
\author[qub]{A.~Alejo}
\author[clf]{S.~Astbury}
\author[york]{C.~Baird}
\author[desy]{S.~Bohlen}
\author[cern]{M.~Campbell}
\author[imp]{C.~Colgan}
\author[cern]{D.~Dannheim}
\author[clf]{C.~Gregory}
\author[hijena,jena]{H.~Harsh}
\author[oxf]{P.~Hatfield} 
\author[umich]{J.~Hinojosa}
\author[hijena,jena]{D.~Hollatz}
\author[clf]{Y.~Katzir}
\author[awe]{J.~Morton}
\author[york]{C.D.~Murphy}
\author[cern]{A.~Nurnberg}
\author[desy]{J.~Osterhoff}
\author[uva]{G.~P\'erez-Callejo}
\author[desy]{K.~P\~{o}der}
\author[clf]{P.P.~Rajeev}
\author[hijena]{C.~Roedel}
\author[jena]{F.~Roeder}
\author[hijena,jena]{F.C.~Salgado}
\author[qub]{G.M.~Samarin}
\author[qub]{G.~Sarri}
\author[hijena,jena]{A.~Seidel}
\author[clf]{C.~Spindloe}
\author[lbnl]{S.~Steinke}
\author[qub,lanc]{M.J.V.~Streeter}
\author[umich,lanc]{A.G.R.~Thomas}
\author[york]{C.~Underwood}
\author[imp]{W.~Wu}
\author[hijena,jena]{M.~Zepf}
\author[imp]{S.J.~Rose}
\author[imp]{S.P.D.~Mangles}

\affiliation[imp]{organization=\IC}
\affiliation[ply]{organization=\PLYMOUTH}

\affiliation[umich]{organization=\UMICH}
\affiliation[desy]{organization=\DESY}
\affiliation[qub]{organization=\QUB}
\affiliation[hijena]{organization=\HIJENA}

\affiliation[jena]{organization=\JENA}
\affiliation[clf]{organization=\CLF}
\affiliation[york]{organization=\YORK}
\affiliation[oxford]{organization=\OXFORD}

\affiliation[awe]{organization=\AWE}
\affiliation[celia]{organization=\CELIA}
\affiliation[uva]{organization=\UVa}
\affiliation[lbnl]{organization=\LBNL}

\affiliation[lanc]{organization=\LANC}
\affiliation[cern]{organization=\CERN}

\vspace{10pt}

\date{\today}

\begin{abstract}
We report on a direct search for elastic photon-photon scattering using x-ray and $\gamma$ photons from a laser-plasma based experiment.
A $\gamma$ photon beam produced by a laser wakefield accelerator provided a broadband $\gamma$ spectrum extending to above \unit[$E_\gamma = 200$]{MeV}. 
These were collided with a dense x-ray field produced by the emission from a laser heated germanium foil at \unit[$E_x \approx 1.4$]{keV}, corresponding to an invariant mass of \unit[$\sqrt{s} = 1.22 \pm 0.22$]{MeV}.
In these asymmetric collisions elastic scattering removes one x-ray and one high-energy $\gamma$ photon and outputs two lower energy $\gamma$ photons.
No changes in the $\gamma$ photon spectrum were observed as a result of the collisions allowing us to place a 95\% upper bound on the cross section of \unit[$1.5 \times 10^{15}$]{µb}.
Although far from the QED prediction, this represents the lowest upper limit obtained so far for \unit[$\sqrt{s} \lesssim 1$]{MeV}. 
\end{abstract}

\maketitle

\section{Introduction}
Photon-photon scattering is one of the most fundamental processes in quantum electrodynamics (QED) and is of elementary importance in astrophysics. 
It is used in models that calculate primordial abundances, affects the observed spectra from $\gamma$-ray bursts from the first million years of the universe \cite{Ellis1992, Svensson:1990pfo} and plays an important role in models of the evolution of strongly magnetised neutron stars \cite{Baring2001}. 
However, these calculations all use the QED cross section which is currently poorly bounded by experiment.

Photon-photon scattering involving virtual photons has previously been observed in several forms (see the summary in Tab.~\ref{tab:diags}): the $1$-to-$1$ process of Delbr\"uck scattering $(\gamma\gamma^* \rightarrow \gamma\gamma^*)$, where a real photon, $\gamma$, scatters from a virtual photon, $\gamma^*$, in the Coulomb field of an ion \cite{Jarlskog:1973aui,Schumacher1975,Akhmadaliev:1998zz}; the $1$-to-$2$ process of photon splitting $(\gamma\gamma^* \rightarrow \gamma\gamma)$ in atomic fields \cite{Jarlskog:1973aui,Akhmadaliev:2001ik}, and the $0$-to-$2$ process of real double photon emission from colliding the virtual photons from Coulomb fields in ultra-peripheral heavy-ion collisions at the ATLAS and CMS experiments $(\gamma^*\gamma^* \rightarrow \gamma\gamma)$ \cite{ATLAS:2017fur,CMS:2018erd,ATLAS:2019azn}.
Instead, in this paper,  we will report on a search for the $2$-to-$2$ process of photon-photon scattering involving only real photons $(\gamma\gamma \rightarrow \gamma\gamma)$.

\begin{table}[h!]
\centering

\resizebox{8.5cm}{!}{\begin{tabular}{ cccc }
\includegraphics[width=0.2\columnwidth]{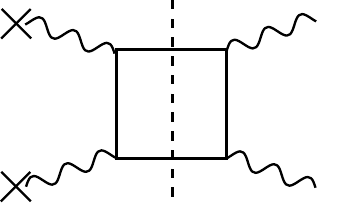}~ &  \includegraphics[width=0.2\columnwidth]{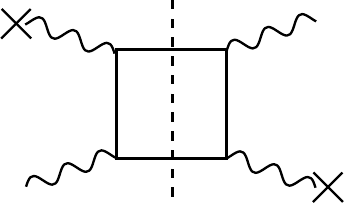}~ &
  \includegraphics[width=0.2\columnwidth]{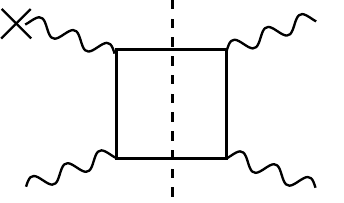} &\!\!\!\!\includegraphics[width=0.2\columnwidth]{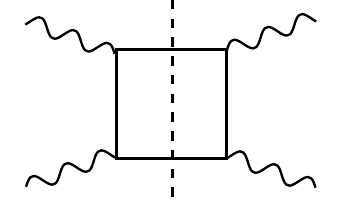} \\ 
 \hline 
  $0$-to-$2$ & $1$-to-$1$ & $1$-to-$2$ & \!\!\!\!$2$-to-$2$ \\
  Landau\cite{Landau:1934zj} & \multicolumn{2}{c}{Bethe-Heitler\cite{Bethe:1934za}}&\!\!\!\! Breit-Wheeler\cite{Breit:1934zz} \\
 \hline
\end{tabular}}

\caption{Comparison of different photon-photon scattering processes. 
The top row shows the Feynman diagram, the second row shows the number of real incoming and outgoing photons ($N-$to$-M$) and the third row shows name of the relevant inelastic contribution (i.e. the real electron-positron pair creation process described by the left-hand side of dotted line in the Feynman diagram). 
In this paper we report results on bounding the cross-section of $2$-to-$2$ scattering.} \label{tab:diags}
\end{table}

A crucial parameter in photon-photon collisions is the invariant mass of the collision $\sqrt{s}$.
For two photons, energy $E_1$ and $E_2$ colliding at an angle $\phi$, we have $s = 2E_1 E_2 (1 - \cos{\phi})$. 
 Photon-photon scattering has been searched for indirectly in the signal of vacuum birefringence at small invariant mass $\sqrt{s}\ll m_ec^2$ in cavity experiments such as PVLAS \cite{Ejlli:2020yhk} and BMV \cite{Agil:2021fiq} involving photons traversing a quasi-constant magnetic field. 
At invariant mass $\sqrt{s}\sim O(\textrm{eV})$ it has also been searched for by directly colliding two optical laser pulses \cite{Moulin:1996vv} and three optical laser pulses \cite{Moulin:1999hwj,Bernard:2000ovj}.
At $\sqrt{s}\sim O(\textrm{keV})$, the cross-section has been bounded by experiments employing x-ray free electron lasers \cite{Inada:2014srv,Yamaji:2016xws}. 
By comparison, for large invariant mass $\sqrt{s} \gg m_{e}c^{2}$, photon scattering with quasi-real photons has been measured by the ATLAS and CMS experiments (in which $\sqrt{s} \approx 5-20\,\textrm{GeV}$). 
Despite these results, photon-photon scattering has yet to be measured using manifestly real photons and this has sustained interest in the process.  
The upcoming HIBEF experiment plans to provide the first measurement involving manifestly real photons at $\sqrt{s}\sim O(10^{2}\,\textrm{eV})$ by colliding an x-ray free electron laser with a high power optical laser pulse \cite{Schlenvoigt:2016jrd, KarbsteinPRL2022} and there have been many suggestions for how to measure this effect using only PW-class optical lasers \cite{Marklund:2006my,DiPiazza:2011tq,King:2015tba,Fedotov:2022ely}.

Apart from being a test of fundamental QED, searches for photon-photon scattering can also provide bounds on physics beyond the Standard Model (BSM),  e.g. the ATLAS results enabled bounds on Born-Infeld electrodynamics \cite{Ellis17} for energy scales $>100\,\textrm{GeV}$. 
Suggestions for improving these bounds by measuring photon-photon scattering at future colliders have also recently appeared in the literature \cite{Ellis:2022uxv}.

 This paper reports on a search for elastic photon-photon scattering at \unit[$\sqrt{s} \approx 1$]{MeV}. This is a $2\to2$ process with two free photons in the out state, in contrast to the $1\to1$ stimulated photon scattering process searched for in cavity experiments (see e.g. \cite{Heinzl:2024cia} for a comparison).
 The search at this energy is motivated at this energy scale because it is relevant to astrophysics, it is where the cross-section takes its maximum value, and at this energy, the QED effect has only very weak experimental bounds. 
 This scale also includes energies over the threshold for creation of real electron-positron pairs via the linear Breit-Wheeler process, which therefore is a sub-process of photon-photon scattering at these energies (this can be understood by the optical theorem \cite{Schwartz_2013}). 
 Linear Breit-Wheeler is also being searched for at PW-class optical lasers \cite{Kettle:2021ipe, ribeyre2016pair}, and multi-photon Breit-Wheeler pair-creation forms part of the science goals for the LUXE experiment planned at DESY \cite{Abramowicz:2021zja,LUXE:2023crk} and the E320 experiment at SLAC \cite{Chen:22}.

Our experiment uses \unit[$\sim 1$]{GeV} electrons from a laser wakefield accelerator \cite{esarey2009physics} which  collide with a fixed target to generate a broadband bremsstrahlung distribution of $\gamma$ photons extending to \unit[$E_\gamma \approx 800$]{MeV}.
The $\gamma$ photons are then collided with the dense x-ray photon field in the vicinity of a laser heated germanium foil.  
The x-ray radiation is dominated by M-L-band emission in the region of  \unit[$E_x \approx 1.4$]{keV}. 
The spectrum of the $\gamma$ photons is determined using a caesium-iodide stack spectrometer \cite{behm2018, Cole2018}. 

\begin{figure}
    \centering
    \includegraphics[width=0.8\columnwidth]{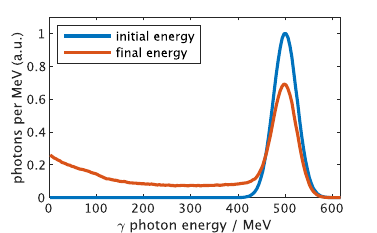}
    \caption{Illustration of the effect on the photon energy spectrum of copious photon-photon scattering of a \unit[500 $\pm$ 25 ]{MeV} $\gamma$ photon beam in a very dense field of \unit[5]{keV} photons with $n_x = 10^{27}$~photons~mm$^{-3}$ and length  \unit[1]{cm}. 
    Scattering in such a high density field results in a broadening and downshift of the $\gamma$ photon energy.} 
    \label{principle}
\end{figure}

In this asymmetric collision, a scattering event typically removes one x-ray and one high energy $\gamma$ photon from the beam and replaces it with two lower energy $\gamma$ photons. 
If sufficient numbers of $\gamma$ photons were to scatter in the x-ray field, we would therefore be able to detect the effect in the photon energy spectrum.
This effect is illustrated in Fig.~\ref{principle} for the idealised case of a \unit[500 $\pm$ 25 ]{MeV} $\gamma$ photon beam  scattering in a very dense field of \unit[5]{keV} photons with $n_x = 10^{27}$~photons~mm$^{-3}$ and length \unit[1]{cm}.
If however, insufficient scattering events occur to produce a measurable  change in the spectrum, this allows us to place a limit on the photon-photon scattering cross-section. 

As well as the effect on the photon spectrum, scattering also leads to an increase in the divergence of the photon beam.
Due to the asymmetry in our photon energies very few photons will be scattered outside of the original $\gamma$ beam profile and so, for our set-up, the effect on the spectrum is more pronounced. 
We therefore use the energy spectrum diagnostic only.

The QED prediction was calculated by extending the Geant4 simulation framework \cite{Geant4, watt2023monte} to include the cross-section for photon-photon scattering \cite{Karplus:1950zz,Karplus:1950zza,DeTollis:1964una,DeTollis:1965vna,Costantini:1971cj}. 
Simulation results were compared to an independent direct numerical evaluation of the process integrated over the x-ray and $\gamma$-distributions reported in experiment,  and found to be in agreement. 
The extended Geant4 model simulates the leading order (in fine-structure constant $\alpha$) photon-photon scattering contribution involving four photons \footnote{During preparation of the manuscript, another simulation framework was developed that includes the same process \cite{Sangal:2021qeg}.}. 
This complements other recently developed simulation frameworks \cite{King:2014vha,Bohl:2015uba,Blinne:2018nbd,Grismayer:2016cqp,Lindner:2022alm} of photon-photon scattering at $\sqrt{s}\sim O(\textrm{eV})-O(\textrm{keV})$ based on the weak-field Heisenberg-Euler Lagrangian.

\section{Experimental set-up}
\begin{figure}
    \centering
    \includegraphics[width=\columnwidth]{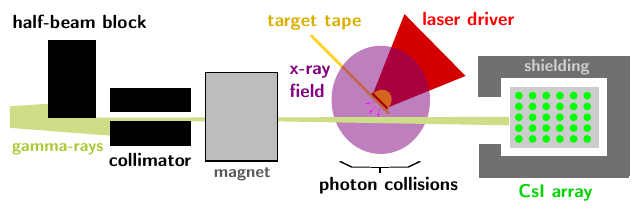}
    \caption{Schematic of the experimental set-up (not to scale). $\gamma$-rays are produced using a laser wakefield accelerator, they are collimated and pass through the x-ray field in the vicinity of a laser-heated plasma. 
    The spectrum of $\gamma$-rays after passage through the x-ray field is characterised using an array of caesium-iodide crystals.} 
    \label{setup}
\end{figure}

The experiment took place at the Gemini laser facility in the UK.
This is a dual beam, \unit[300]{TW} Ti:Sa system, allowing us to generate and collide two high energy-density photon sources. 
The experimental setup was based on the scheme by Pike \emph{et al.,} (2014) with asymmetrical photon sources \cite{Pike}. 
The $\gamma$ photon source was provided by bremsstrahlung emission produced by an electron beam in a high Z-material target. 
The electron beam was produced using laser wakefield acceleration\cite{tajimadawson1979}.
The x-ray photon source was generated through direct laser heating of a thin metal foil. 
The two photon beams were temporally overlapped at the interaction point within 2 picoseconds using the drive laser beams (same optical path) and a fast-response photodiode.
A schematic of the  experimental setup can be found in figure 
\ref{setup}.
A detailed description of our  laser-plasma platform for photon-photon physics can be found in ref \cite{Kettle:2021ipe}.
The experiment can be separated into three parts: the x-ray photon source, the $\gamma$ photon source, and the $\gamma$ photon spectrometer

\subsection{x-ray photon source}
One of the Gemini laser pulses was used to generate a dense x-ray field by rapidly heating a \unit[100]{nm} germanium (Ge) foil. 
As this solid Ge foil is heated, it turns into a plasma, leading to the emission of intense x-ray radiation predominantly due to M-L band transitions \cite{bearden1967, Kettle2015}. 

The heating laser pulse had a duration of \unit[40]{ps}  (\textsc{fwhm} intensity) and a total energy of  \unit[$10.7 \pm 0.3$]{J}. 
It was focused to an elliptical spot using a distributive phase plate, with major and minor axes of \unit[$(217 \pm 6)$]{{\textmu}m} and \unit[$(77 \pm 6)$]{{\textmu}m} respectively, which contained 72\% of the total energy.
The Ge targets were mounted on a Kapton (C$_{22}$H$_{10}$N$_2$O$_5$) tape with a lower average atomic number, limiting the mass of Ge close to the interaction which is a potential noise source.
A motorised tape-drive was used to change targets between shots.

To diagnose the x-ray field, a pinhole imaging system and crystal spectrometer were used. 
The pinhole imaging system gave an on-shot measure of both the emission spot size and the target alignment. 
The spectrometer used a flat, thallium acid phthalate (TlAP) crystal, with a spectral window of \unit[$\approx 700$]{eV}, centred at approximately \unit[$1.6$]{keV} (although the signal above approximately \unit[$1.5$]{keV} was supressed due to an aluminium filter). 
This spectral window is around the M-L band transitions of Ge.
A measurement of the x-ray spectrum, averaged over 47 shots is shown in fig \ref{photondata}a).

The x-ray spectrometer used to measure the x-ray spectrum was absolutely calibrated by taking into account transmission through filters, camera sensitivity and crystal reflectivity. 
At a distance from the source larger than a few times the source size (as is the case at the collision point), the emission can be treated as coming from a spherically symmetric point source. 
Knowledge of the solid angle captured by the spectrometer allows us to calculate the x-ray photon density and extent at the collision point. 
The measured total conversion efficiency from laser energy to 1.3–1.5 keV x-rays was ($2.4 \pm 0.3)\%$. 
This corresponds to \unit[$(3.7\pm0.4)\times 10^{10}$]{photons\,eV$^{-1}$\,J$^{-1}$\,srad$^{-1}$} emitted normal to the front surface of the germanium target.
Taking into account the absorption in the kapton layer on the rear side of the target, at the interaction region (1\,mm from the tape) this corresponds to an x-ray photon density of \unit[$(1.4 \pm 0.5) \times 10^{12}$]{mm$^{-3}$}, over an effective length of approximately \unit[3]{mm}. 

\subsection{$\gamma$ photon source}
 
The $\gamma$ photon source was generated through bremsstrahlung emission, which first requires a beam of high energy electrons. 
To produce these electrons, one of the Gemini laser beams was focused into a \unit[17.5]{mm} gas cell filled with helium and a 2\% nitrogen dopant. 
The duration of the laser pulse was \unit[$45 \pm 5$]{fs} (\textsc{fwhm} intensity) and the focal spot was \unit[$(44 \pm 2)$]{{\textmu}m} $\times$ \unit[$(53 \pm 2)$]{{\textmu}m} (\textsc{fwhm} intensity). 
The laser  energy on target was $5.5 \pm 0.6$ J, corresponding to a normalised vector potential $a_0 = 1.1 \pm 0.2$.
Through the laser wakefield acceleration mechanism \cite{esarey2009physics}, a beam of high energy electrons (energy up to \unit[$\approx 800$]{MeV}, charge \unit[$\approx 50$]{pC} ) was emitted from the gas cell.

These electrons then passed through a \unit[0.5]{mm} thick bismuth (Bi) foil, acting as a bremsstrahlung converter. 
This emits a beam of high energy $\gamma$ photons with a similar duration to that of the driving laser pulse (i.e $\sim 50$~fs).

A calculation of the $\gamma$ photon spectrum produced in the experiment,  based on Geant4 calculations of the bremsstrahlung conversion process for the measured electron beam spectrum is shown in figure \ref{photondata}b).

\subsection{Spatio-temporal alignment}

To achieve successful collisions between the two photon sources, it is necessary to overlap the sources in both space and time with sufficient precision.  
One of the advantages of this experiment is that the precision required to achieve collisions is not difficult to achieve. 
As the photon sources are driven by optical lasers it is straightforward to overlap the two sources with \unit[1]{ps} and \unit[10]{\textmu m} precision by overlapping the drive laser pulses using fast diodes and optical imaging. 
Since the x-ray field size is $\sim 1\,$mm and the x-ray field duration is $\approx 40\,$ps, this is more than sufficient to achieve spatial and temporal overlap of the photon sources at the desired collision location. 

\subsection{Background}

The bremsstrahlung process used to generate the high-energy $\gamma$ rays also generates a large number of low energy, divergent $\gamma$ photons, which are a potential noise source in this experiment. 
If these divergent $\gamma$ photons were to interact with the Ge foil, or another part of the experimental setup, they would produce background through the Compton scattering process. 
To prevent this a \unit[100]{mm} block of tungsten (W) with a \unit[2]{mm} diameter hole drilled through the centre was used to collimate the $\gamma$ photon beam.
This collimator effectively removes any $\gamma$ photon traveling at an angle greater than \unit[10]{mrad} to the beam propagation direction. 

To further reduce the background, a \unit[50]{mm} tungsten block was placed just off-axis, shadowing the Ge foil from the $\gamma$ photon beam. 
Placing such high Z-material close the $\gamma$ photon beam axis will itself generate a large number of background Bethe–Heitler pairs which would pass through the x-ray field and create a background signal through Compton scattering. 
Therefore, a \unit[30]{cm} dipole magnet with a field strength of \unit[$B = 1$]{T}, was used to remove these before the photon-photon interaction zone.

The target design and experimental geometry ensure that no plasma is in the path of the $\gamma$ ray beam when the $\gamma$ rays pass near the x-ray generating foil.  
This is achieved by mounting the germanium foil on a $\unit[4]{\mu m}$ kapton layer.  
The x-rays pass through the kapton to reach the collision point but the kapton layer prevents the expansion of any plasma towards the collision volume.  
Even without this kapton layer plasma expansion would not be an issue as the closest that the edge of the collimated $\gamma$ beam passes to the foil is \unit[1]{mm}, \unit[40]{ps} after the start of the laser pulse. Laser produced plasmas have ablation speeds of \unit[100 – 10000]{km~s$^{-1}$}, so any plasma would only have expanded by \unit[0.1 – 1]{$\mu$m} at the time of the arrival of the  $\gamma$ photons. 

Any effect of $\gamma$ photons interacting with material in the chamber, including the x-ray target foil are captured in the analysis by comparing the $\gamma$ spectrum on collision shots with null shots where the experimental geometry is identical but there is no laser incident on the x-ray target foil. 

\subsection{$\gamma$ photon spectrometer}

The $\gamma$ photon spectrometer consisted of an array of  $5 \times 5 \time 50$~mm crystals of CsI doped with thallium.
The array was arranged in 47 columns ($z$) and 33 rows ($y$), with the long side of the crystals ($x$) oriented transversely to the propagation direction of the $\gamma$ rays ($z$).  
The deposition of energy by the incident $\gamma$-rays inside the array was captured by imaging the light emission by the crystals with an EMCCD camera. 
The deposition in the $z$-direction can be used to infer the $\gamma$-ray energy spectrum, while the  response in the $y$-axis encodes information about the vertical divergence of the radiation.
For our analysis we integrate the signal along the $y$-axis, treating each column of CsI crystals together. 
The spectrometer method is described in more detail in \cite{Kettle:2021ipe,behm2018}.

The $\gamma$-photon spectrum is determined using a forward model based on a trial function.
Bayesian inference is used to determine the best-fit parameters of the trial function and their uncertainty.

The detector was calibrated to remove systematic effects such as variations in light yield in each crystal due to crystal imperfections and misalignment, or effects produced by the imaging system.
We compared the measured energy deposited in each crystal averaged over a large number of shots with that predicted in Geant4 using the average electron energy spectrum (as measured on a series of shots without the bismuth foil intercepting the electron beam) and generated a correction factor which can then be applied to each column of crystals.  

By running Geant4 simulations for a series of mono-energetic $\gamma$ photon beams over a range of energies we can model response of each  crystal as a function of photon energy, $\rho_i(E_\gamma)$, where $i$ is the crystal index and $E_\gamma$ is the photon energy.
From this the signal measured by the detector for an arbitrary $\gamma$ photon spectrum can be quickly calculated with the following integral
\begin{equation}
 I_{i}[f]=C_{i} \int_{0}^{\infty} \rho_{i}(E) f(E) \mathrm{d} E
 \label{signal equation}
\end{equation}
where $C_i$ is the correction factor.

To enable a Bayesian inference of the $\gamma$ photon spectrum on each shot, we first find a low dimensional parameterisation of the spectrum,  f(E). 
We performed Geant4 simulations of the bremsstrahlung converter, using electron energy spectra observed on the experiment, to get a data set of typical $\gamma$ photon spectra. 
From this data set, the following function was found to provide a good approximation to the spectra measured on the detector.
\begin{equation}
f\left(E ; \alpha, E_{c}\right)=\alpha\left(1-0.18^2 \frac{E^{2}}{E_{c}^{2}}\right) E^{-0.94} \,,
\end{equation}
where $\alpha$ controls the amplitude of the spectrum and $E_{c}$ is a characteristic energy that controls both the slope and the position of the cut off at high energy.
Using this parameterisation, we can  obtain the $\gamma$ photon spectrum on each shot by applying Bayesian inference to estimate a distribution over $\mathbf{x}=\left(E_{c}, \alpha\right)$.
This involves applying Bayes’ theorem
\begin{equation}
p(\mathbf{x} \mid \mathbf{y})=\frac{p(\mathbf{y} \mid \mathbf{x}) p(\mathbf{x})}{\int p(\mathbf{y} \mid \mathbf{x}) p(\mathbf{x}) \mathrm{d} \mathbf{x}}\,, 
\label{Bayes}
\end{equation}
where $\mathbf{y}$ is an observed data point, corresponding to a vector of the crystal responses. 
In this equation $p(\mathbf{y} \mid \mathbf{x})$ is the likelihood and $p(\mathbf{x})=p\left(E_{c}\right) p(\alpha)$ is a prior which we must set.
If we make the assumption that the crystals exhibit random Gaussian noise, $\sigma$, we can
 write the likelihood function as
 $$ p(\mathbf{y} \mid \mathbf{x})=\prod_{i} \frac{1}{\sigma \sqrt{2 \pi}} \exp \left(-\frac{\left(y_{i}-I_{i}[f(\mathbf{x})]\right)^{2}}{\sigma^{2}}\right) \,.$$
 This introduces a new parameter, $\sigma$, which is treated in the same way as $E_c$ and $\alpha$. 
 Given that we have little prior knowledge of $E_c$, $\alpha$ or $\sigma$, other than the fact that they cannot be negative, we  set uniform priors on each with a lower bound of zero.
 We know the upper bound for $E_c$ cannot be greater than the maximum energy of the electrons (\unit[$\approx 800$]{MeV}) so the prior used is $p\left(E_{c}\right)=\mathcal{U}(0,800 \mathrm{MeV})$.
 Through appropriate normalisation of the data set, we can ensure that $\alpha$ is never greater than 10, allowing us to apply the prior $p(\alpha)=\mathcal{U}(0,10))$. 
 Finally, we  set the prior on $\sigma$ to be $p(\sigma)=\mathcal{U}(0,1)$ as if the limit is greater than this, the data will be too noisy to make any inference.
 
\begin{figure}
   \centering
   \includegraphics[width=\columnwidth]{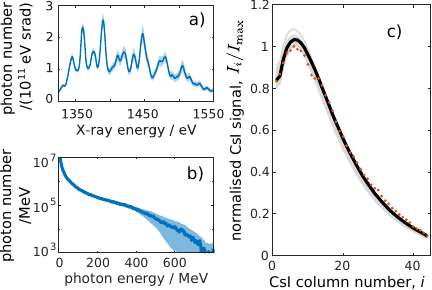}
   \caption{a) Measured x-ray photon spectrum, showing average and standard deviation measured over 47 shots, 
   b) $\gamma$ photon spectrum (Geant4 simulations based on measured electron beam, showing average and standard deviation measured over 10 shots)  c) Measured signal recorded by the CsI array from a randomly selected null shot (red circles). 
   The signal is the intensity of light emitted by each column of CsI crystals (i.e eqn \ref{signal equation}), normalised to the peak signal.  
   The signal generated by the forward model of the detector for the  average  (black) and 100 randomly selected samples (grey) samples from the posterior distribution of $E_c$ is also shown. 
   The mean value corresponds to \unit[$E_c = 83.2 \pm 3.2 $]{MeV}.}
   \label{photondata}
\end{figure}

 With the likelihood and priors set, we can use equation \ref{Bayes} to calculate the posterior. 
 Given that the numerator involves a three-dimensional integral, it is most efficiently solved using a Markov chain Monte Carlo (MCMC) method. 
 In figure \ref{photondata}c) we can see an example of this calculation performed on a randomly selected shot from the data set.

 \section{Results}
\subsection{$\gamma$ photon spectrum }
Having developed a robust method for extracting the $\gamma$ photon spectrum from the crystal response, we can test if the presence of the x-ray field has an effect on the $\gamma$ photon spectrum. 
To do this, we  run the Bayesian spectral retrieval algorithm on each shot of the experiment and compare the distributions over $E_c$ for null and collision shots.
Null shots involve firing only the beam that generates the $\gamma$ photon beam. 
Collision shots involve firing both beams at a relative delay that ensures the $\gamma$ photons pass through the x-ray field.
The Ge foil was properly aligned for both null and collision shots to ensure that any contribution to the $\gamma$ photon spectrum measurement due to interactions with the foil are fully accounted for.
The data set consists of 32 null shots and 22 collision shots. 
These shots were all performed on a single shot day on the Gemini laser system.

\begin{figure}
    \centering
    \includegraphics[width=8.5cm]{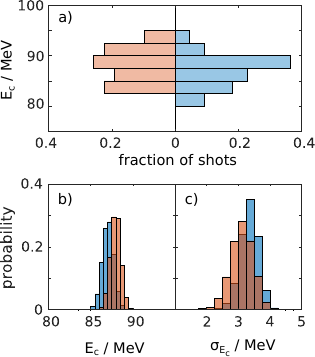}
    \caption{a) Distribution of inferred value for the $\gamma$ photon spectrum $E_c$  for 32 null shots (blue) and 22 collision shots (red) 
    b) Bootstrap estimate of the mean of $E_c$ for null shots (blue) and collision shots (red)
    c) Bootstrap estimate of the standard deviation of $E_c$ for null shots (blue) and collision shots (red)}
    \label{stats}
\end{figure}

We compare the distribution of $E_c$ on collision and null shots in various ways.
Figure \ref{stats}a) shows the histogram of the inferred value for the $\gamma$ photon spectrum $E_c$. 
The relatively small number of shots in each distribution means it is not immediately clear if  differences between them are significant. 

Figure \ref{stats}b) shows the empirical cumulative distribution function (ecdf) of $E_c$  for the data.  
Also shown are 50 ecdfs for bootstrap samples of the data, these effectively represent the uncertainty in the ecdfs. 
The overlap between these ecdfs illustrates that there is no significant difference between the distributions.
Figures \ref{stats}c) and d) show the distribution of mean and standard deviation calculated from 100,000 bootstrap samples of the null and collision shot data.  
The fact that the distributions overlap  further illustrates that there is no significant difference between the distribution of $E_c$ on null and collision shots. 

A third method to assess differences in the distribution of $E_c$ is to use the  two-sample Kolmogorov–Smirnov (KS) test. 
The null hypothesis of this test is that both null and full data sets have been sampled from the same distribution, i.e. that there is no measurable difference effect of the photon-photon collisions on the measured $\gamma$-ray spectrum. 
To perform the KS test, the two-sample KS-test statistic must be calculated:
\begin{equation}
    D=\sup \left|F_{\mathrm{N}}\left(E_{c}\right)-F_{\mathrm{C}}\left(E_{c}\right)\right| \,,
\end{equation}
where $F_{\mathrm{N}}\left(E_{c}\right)$ and $F_{\mathrm{C}}\left(E_{c}\right)$  are the cumulative distribution functions for the null and collision shots respectively. 
The null hypothesis is accepted at the 95\% confidence level if $D < 0.378$\cite{knuth} for data sets with 32 and 22 samples. 
The value obtained for our data set is D = 0.216, so we cannot reject the null hypothesis.

\begin{figure}
     \centering
     \includegraphics[width=\columnwidth]{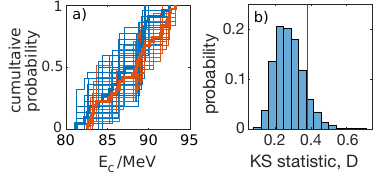}
     \caption{a) The empirical cumulative distribution (ecdf) function of $E_c$ for null shots (blue) and collision shots (red). 
    The thick lines show the ecdf for the experimental data, the thin lines show ecdfs calculated from 50 bootstrap samples of the data.
    b) Distribution of the two sample Kolmogorov-Smirnov test statistic comparing the distribution of $E_c$ for null shots and collision shots from 100,000 bootstrap samples from the data. 
     The red line shows the value above which the null hypothesis (that the null and collision shots are  from the same distribution) can be rejected at the $\alpha = 0.05$ significance level.  }
     \label{kstest}
\end{figure}

We can also calculate the two-sample Kolmogorov-Smirnov test statistic for a large number of bootstrap samples from the data to estimate the uncertainty in the KS test statistic (shown in Fig. \ref{kstest}). 
We find that the bulk of the distribution ($\approx 90\%$) lies below the critical value, providing further strong evidence that we cannot reject the null hypothesis, i.e. we must assume that collisions between $\gamma$ photons and the dense x-ray field  did not produce a detectable difference in the energy spectrum of the $\gamma$ photons. 

\subsection{Bounding the cross section}
As the various analyses all show that there  is no significant difference between the distribution over $E_c$ on null and collision shots, we can conclude there was not a detectable level of photon-photon scattering.
To find how much larger than the standard QED value the cross-section would have to be to produce a detectable, we  performed multiple simulations of the experiment with an increasing cross-section. 
The factor by which the cross section would have to increase for us to have observed a significant difference in the value of $E_c$ on collision shots provides a bound on the cross-section.
These simulations were performed with Geant4 \cite{Geant4} which we have adapted to include photon-photon collisions between the $\gamma$ photons and a dense x-ray field, using the QED cross section \cite{watt2023monte}.

The simulation modelled all major aspects of the experimental geometry, including the $\gamma$-photon source, collimator and magnet before the collision point, the x-ray photon source (both the tape target and surrounding x-ray field), and the magnetic transport system, shielding and detectors and after the collision point. 
The x-ray photon source was modelled as a static (i.e. non evolving) photon field  using the measured x-ray spectrum and a spatial distribution of photon density $n_x(x,y,z)$ calculated from the experimentally measured photon numbers and assuming emission from a uniform disk of radius \unit[100]{nm}. 
Details of the geometry can be found in \cite{Kettle:2021ipe}, and details of the modifications made to Geant4 are described in \cite{watt2023monte}.

\begin{figure}
    \centering
    \includegraphics[width=\columnwidth]{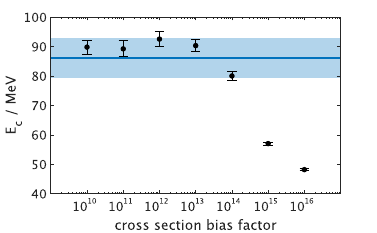}
    \caption{Simulated $\gamma$ photon spectrum critical energy, $E_c$, after interaction with the x-ray field, as a function of the cross section bias factor. 
    Blue shaded region represents the critical energy measured on null shots (95\% confidence interval).}
    \label{bias_factor}
\end{figure}

The result of these simulations is shown in figure \ref{bias_factor}.  
Also shown is the mean and 95\% confidence limit of $E_c$ for the null shots. 
Increasing the bias on the cross-section up to $10^{13}$ has little effect on the $E_c$ our detector would measure.
Beyond this point, $E_c$ starts to decrease.
The copious amount of elastic photon-photon scattering that would occur if the cross section were $10^{14} - 10^{15}$ times higher than the QED prediction would significantly lower the average energy of the $\gamma$ photons exiting the collision volume. 
The simulations show that this would result in a measurably lower value of $E_c$ on collision shots than that measured on null shots.
The fact that we do not measure a lower value of $E_c$ on collision shots therefore allows us to place an upper bound on the elastic photon-photon scattering cross section at $\approx 10^{15} \sigma_{\rm QED}$. 

The broadband nature of the photon spectra in this experiment means that this measurement is not at a single, specified value of $\sqrt{s}$, but the effective $\sqrt{s}$ be found by weighting the cross section, $\sigma(E_1, E_2)$, with the measured photon spectra $N_x(E_1)$ and $N_\gamma(E_2)$ and considering the range of collision angles.
The effective $\sqrt{s}$ for this experiment is \unit[$1.22 \pm 0.22$]{MeV}.

\section{Conclusions}

\begin{figure}[h!!]
    \centering
    \includegraphics[width=0.9\columnwidth]{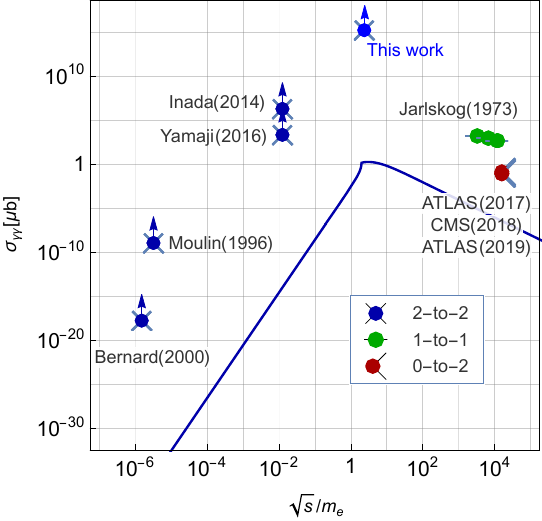}
    \caption{Comparison of this work with different measurements of the total cross-section of photon-photon scattering. 
    The theory prediction for the cross-section of the $2$-to-$2$ process is plotted in the solid blue line (calculated from \cite{DeTollis:1964una,DeTollis:1965vna,Costantini:1971cj}).  For the $1$-to-$1$ process (also measured in \cite{Akhmadaliev:1998zz}), the horizontal co-ordinate is given by the incident photon energy in the lab. For the $0$-to-$2$ process, the measurements were for a diphoton mass $> 5\,\textrm{GeV}$ \cite{CMS:2018erd} or $> 6\,\textrm{GeV}$ \cite{ATLAS:2017fur,ATLAS:2019azn}, to represent this on the plot, $\sqrt{s}=10\,\textrm{GeV}$ has been chosen.
    Notable exceptions to this plotting scheme are the cavity experiments such as PVLAS \cite{Ejlli:2020yhk} and BMV \cite{Agil:2021fiq}, measuring a $1$-to-$1$ process in a quasi-constant magnetic field and photon-splitting experiments \cite{Jarlskog:1973aui,Akhmadaliev:2001ik} measuring a $1$-to-$2$ process.}
    \label{cross_section_comp}
\end{figure}

The cross-section limits that have been made by previous direct searches for elastic photon-photon scattering are shown in figure \ref{cross_section_comp}.
The closest of these to the QED cross-section for real photon-photon scattering is that of Bernard et al. (2000) \cite{Bernard:2000ovj} using optical photons. 
However, this is a factor of $10^{18}$ times higher than the QED prediction.
More recently, work using x-ray photons provided by a free electron laser bounded the cross section at $\sqrt{s} \sim 10^{-2} m_e$ a factor of $10^{19}$ times higher than the QED prediction \cite{Inada:2014srv, Yamaji:2016xws}. 
These high bounds are due to the fact that these previous direct searches operated in a regime where $\sqrt{s} \ll m_e$, where the cross-section is severely suppressed.
The experiment reported here provides the first bound at $\sqrt{s} \sim m_e$, where elastic scattering is expected to play a role in various astrophysical situations \cite{Svensson:1990pfo}.
This experiment also provides the lowest ratio of the upper bound to the QED prediction for $2$-to-$2$ photon-photon scattering to date and the lowest bound in the range close to \unit[$\sqrt{s} \approx 1$]{MeV}.

While this current work provides an upper bound on the cross section, it is also useful to consider if laser-plasma interactions are a potential route to directly observing photon-photon collisions in the laboratory. 
To do this we consider how long an experiment would have to operate to observe a single scatter event. 
A simple estimate of the number of scatter events per shot is
$ N_{\rm scatter} \approx N_\gamma \sigma n_x L_x$, where $N_\gamma$ is the number of $\gamma$ photons, $\sigma$ is the cross section, $n_x$ is the x-ray photon density and $L_x$ is the length of the x-ray field.  
For the current configuration described here $N_\gamma \sim 10^7$, $\sigma \sim 10^{-30}~{\rm cm}^{2}$, $n_x \sim 10^{15}~{\rm cm}^{-3}$, and $L_x \sim 0.1~{\rm cm}$, resulting in $N_{\rm scatter} \sim 10^{-9}$ per laser shot.  
At the repetition rate of this experiment (0.05~Hz) this would require over 600 years of continuous operation, but a 100 Hz laser with similar capabilities would require only 100 days. 

Higher energy lasers such as EPAC \cite{mason2023EPAC} and ELI-NP\cite{Lureau_ELINP2020} will be capable of producing greater than 10 times more $\gamma$ photons per shot due to the higher charge, higher energy electron beams they will be capable of producing. 
If such lasers could be operated at 100~Hz, the required time drops to $\sim 1$ day.  

The development of such high repetition rate, high power lasers has already been identified as a key future direction for laser wakefield accelerators \cite{Albert-plasmaroadmap2020} and is an area of active research (see e.g. \cite{DeVidoDipole100Hz}).
Such facilities could open up the real possibility of observing and studying photon-photon scattering.

\section*{Data Availability}
The data that support the findings of this study are openly available at the following URL/DOI: \href{https://dx.doi.org/10.5281/zenodo.13767363}{https://dx.doi.org/10.5281/zenodo.13767363}.

\section*{Acknowledgement}
We wish to acknowledge the support of the staff at the Central Laser Facility. 
This project has received funding from the European Research Council (ERC) under the European Union’s Horizon 2020 research and innovation programme (Grant Agreement No. 682399) and STFC (Grant No. ST/P002021/1). 
JH and AT acknowledge support from the US National Science Foundation grant \#1804463. GS would like to acknowledge support from EPSRC (Grant Nos. EP/N027175/1, EP/P010059/1).
GPC was supported by Research Grant No. PID2022-137632OB-I00 from the Spanish Ministry of Science and Innovation.
\bibliography{citations}

\end{document}